\documentclass{article}
\usepackage{graphicx} 
\usepackage[a4paper, left=1in, right=1in, top=1in, bottom=1in]{geometry}
\usepackage[colorlinks=true, linkcolor=blue, citecolor=blue, urlcolor=blue]{hyperref}
\usepackage{natbib}
\usepackage{authblk}
\usepackage{svg}
\usepackage{float}
\usepackage{caption}
\usepackage{graphicx}
\usepackage{amsmath}
\usepackage{amssymb}

\title{An Uncertainty Principle for Probabilistic Computation in the Retina}
\author{Jayanth R Taranath (J.R.T) \& Salim M'Jahad}

\affil{UT Austin \& Saphe Labs}

\begin{document}

\maketitle

\begin{abstract}
We introduce a probabilistic model of early visual processing, beginning with the interaction between a light wavefront and the retina. We argue that perception originates not with deterministic transduction, but with probabilistic threshold crossings shaped by quantum photon arrival statistics and biological variability. We formalize this with an uncertainty relation, \( \Delta \alpha \cdot \Delta t \geq \eta \), through the transformation of light into symbolic neural code through the layered retinal architecture. Our model is supported by previous experimental results, which show intrinsic variability in retinal responses even under fixed stimuli. We contrast this with a classical null hypothesis of deterministic encoding and propose experiments to further test our uncertainty relation. By re-framing the retina as a probabilistic measurement device, we lay the foundation for future models of cortical dynamics rooted in quantum-like computation. We are not claiming that the brain could be working as a quantum-system, but rather putting forth the argument that the brain as a classical system could still implement quantum-inspired computations. We define quantum-inspired computation as a scheme that includes both probabilistic and time-sensitive computation, clearly separating it from classically implementable probabilistic systems. 
\end{abstract}

\section{Introduction}

What does it mean for a biological system to perceive? Standard theories of neural computation often begin with the assumption that a stimulus is given — spatially and temporally — and proceed to examine how that stimulus is encoded, transformed, and interpreted by the brain. But this assumption leaves unaddressed a foundational question: how and when is a stimulus first received and recognized as real within the system? Is the beginning of perception — the moment when light first interacts with the retina — a deterministic event? Or is it intrinsically probabilistic, governed by fundamental constraints arising from the physical characteristics of light and biological variability?

In this manuscript, we take a first-principles approach to visual perception, starting not with cognition or computation, but with the earliest physical interaction between photons and retinal tissue. We propose that the act of perception originates at a quantum-classical boundary: photons, as quantum entities, interact with molecular photoreceptors in the retina via probabilistic absorption events. These events initiate a cascade of transduction and transformation that culminate in the firing of retinal ganglion cells. Importantly, this cascade is not deterministic. Instead, it is shaped by two fundamental sources of uncertainty: (1) the probabilistic arrival and absorption of photons, and (2) the biological variability in neuronal thresholds and response latencies.

To formalize this view, we introduce a quantum-inspired uncertainty principle for perception: \(\Delta \alpha \cdot \Delta t \geq \eta\), where \(\Delta \alpha\) represents the variability in activation thresholds of retinal cells and \(\Delta t\) represents the temporal uncertainty in the generation of neural responses. This formulation parallels the Heisenberg uncertainty principle in quantum mechanics, but is grounded in biophysical and neurophysiological realities.

We examine this uncertainty principle through simulation and analysis of the retinal architecture, stopping just before the information reaches downstream cortical areas. Our focus is on the transformation of a simple, monochromatic spherical light wavefront as it interacts with the retina, and how this interaction leads to the emergence of probabilistic neural representations. We show that even in an idealized, noise-minimized setup, the system exhibits irreducible uncertainty in both the spatial and temporal domains of perception.

To support our theoretical framework, we engage in a detailed comparison with experimental findings, particularly from the work of Baylor and Rieke, who have shown that retinal rods can respond to single photons yet still exhibit nontrivial trial-to-trial variability~\citep{baylor1979responses, rieke1998origin, rieke1998single}. These findings strongly support our claim that perception begins not with a deterministic measurement, but with a probabilistic event shaped by both physical and biological constraints.

Ultimately, this manuscript lays the foundation for a broader theoretical framework in which perception — and potentially cognition — is viewed through the lens of probabilistic state transitions rather than deterministic signal processing. Future work in this series will extend the model to cortical dynamics, exploring the role of oscillations, inter-regional synchronization, and higher-order inference in the brain’s probabilistic construction of the world.

\section{The Architecture of the Visual Stream}

Understanding the probabilistic nature of perception requires a clear view of the biological structure through which sensory signals first pass. In this section, we provide an anatomical and functional overview of the visual stream, with emphasis on the early stages where our theoretical framework applies. We restrict our analysis to the flow of information from the photoreceptors in the retina to the retinal ganglion cells (RGCs), stopping before the optic tract and thalamic relay.

The primate retina is a layered, highly specialized structure composed of several distinct cell types. Light first interacts with rods and cones, which transduce incoming photons into graded electrical signals ~\citep{lagnado1992signal}. These are then relayed to bipolar cells, modulated by lateral input from horizontal cells, and shaped temporally by amacrine cells. The resulting signals converge on the ganglion cells, which fire discrete action potentials that encode the earliest symbolic representation of the visual world.

This architecture is not linear; it is dynamic, recursive, and adaptive~\citep{balasubramanian2009receptive}. Each cell type applies a transformation that reflects both its own biophysical constraints and its place in a broader inferential network. Horizontal cells mediate local contrast through lateral inhibition~\citep{xin1999dark}. Bipolar cells implement gain control and polarity selectivity~\citep{euler2000light}. Amacrine cells impose temporal filters, anticipation, and motion sensitivity~\citep{o1993responses}. Retinal ganglion cells pool these pre-processed inputs into sparse spike trains that serve as the output of this probabilistic front-end~\citep{purpura1990light}.

The structure of the retina — in both morphology and function — reflects evolutionary pressures to extract maximal information under conditions of limited certainty~\citep{lamb2007evolution}. The center-surround organization of receptive fields, the high cone density in the fovea, and the motion-predictive properties of interneurons all support the idea that the retina is not merely a passive relay but a computationally active structure designed for real-time probabilistic inference.

Figure 1 provides a schematic overview of the visual stream modules addressed in this manuscript, with emphasis on the transformations occurring within the retina. It contextualizes the formalism that follows.

\begin{figure}[H]
\centering
\includegraphics[width=0.9\textwidth]{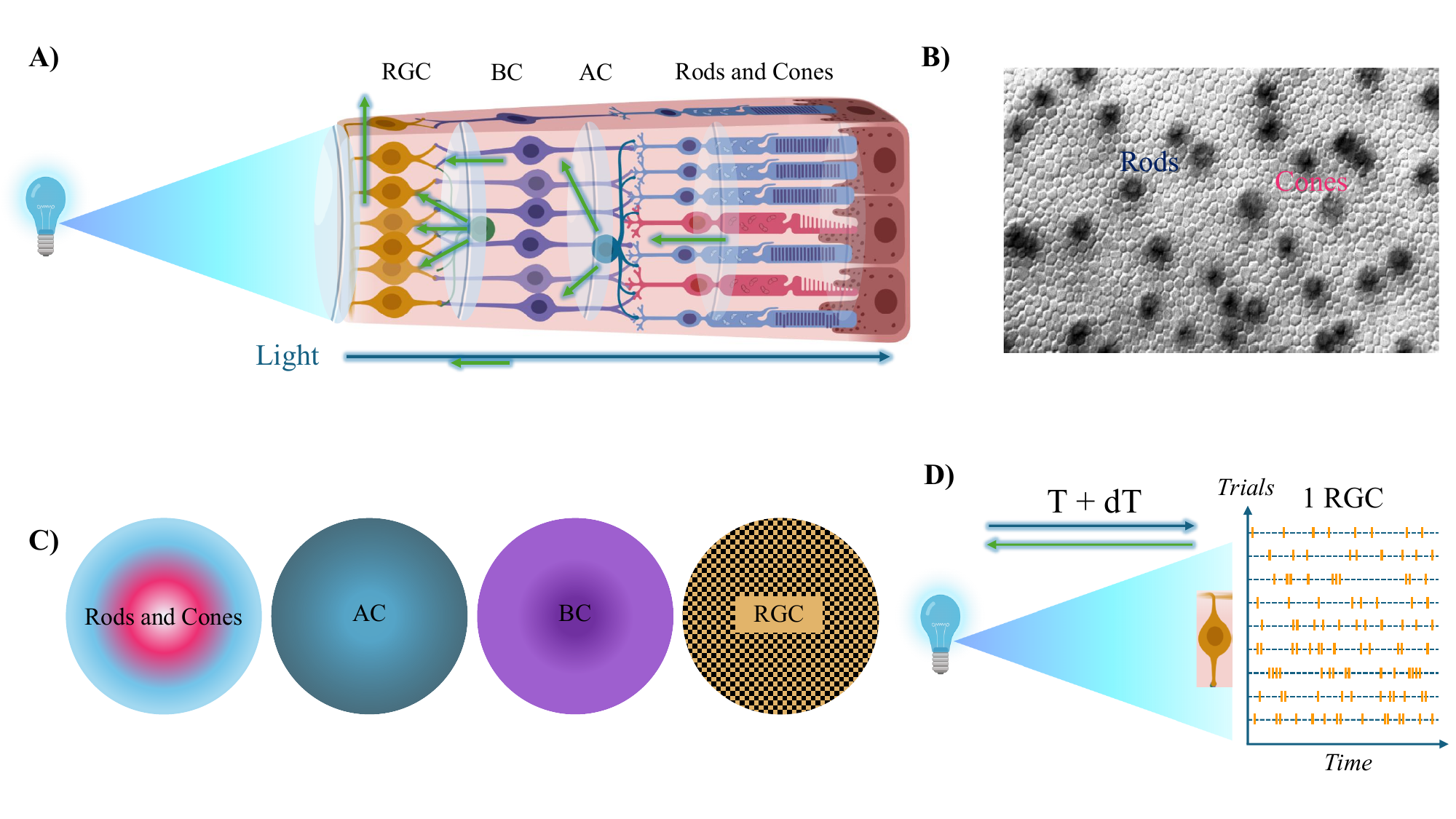} 
\caption{
\textbf{Perception’s First Act — Clouds.} \\
\textbf{A)} A monochromatic light wavefront enters from the left and interacts with the retinal sheet. Rods and cones transduce the light, passing graded signals through bipolar (BC), horizontal cells (not emphasized but assumed inside the formalism containing BC), and amacrine cells (AC) before discrete spike codes emerge at the level of retinal ganglion cells (RGC). Green arrows denote the direction of information flow, with local lateral interactions shown within the retinal layers. \\
\textbf{B)} Micro-anatomical structure of the retina emphasizing how rods and cones are distributed in a non-uniform, hexagonally approximated grid. \\
\textbf{C)} A grid of photoreceptors activated with varying probabilities over time. The color scale reflects the likelihood of activation, which depends on both wavefront intensity and intrinsic threshold variability. The spatiotemporal spread of activation illustrates the uncertainty in identifying the precise moment and location of first perception. We assume that the light wavefront propagates spherically, with intensity falling off radially from the center. RGC is the layer where discrete spikes occur. \\
\textbf{D)} The trial-trial variability in the spike-raster for an example RGC is shown with the presentation of the same stimulus (intensity, shape, size, orientation, and distance)~\citep{berry1997structure}.
}
\end{figure}

\section{Microscopic Structure of the Retina} 

To fully grasp how the retina performs probabilistic inference on incoming light stimuli, we must look beyond gross anatomical flow and examine the cellular microstructure and spatial organization of retinal components. Each class of neuron within the retina contributes uniquely to the transformation of quantum-scale input signals into usable neural code.

Photoreceptors — rods and cones — form a semi-regular, quasi-hexagonal lattice across the retina~\citep{steinberg1973distribution}. This arrangement supports high spatial sampling density while allowing for locally variable cell densities. Rods dominate the peripheral retina and are optimized for sensitivity in low-light conditions, while cones are concentrated in the fovea and support high-acuity, color-sensitive vision~\citep{rieke1998origin}. This spatial heterogeneity already encodes an evolutionary solution to the problem of probabilistic light sampling: greater certainty is allocated where visual information is densest.

These photoreceptors synapse onto bipolar cells, whose receptive fields are organized in a center-surround topology, sculpted by inhibitory input from horizontal cells~\citep{mcguire1984microcircuitry}. The center-surround motif serves to enhance local contrast and suppress redundant input — a transformation consistent with early stages of Bayesian inference~\citep{balasubramanian2009receptive}. Amacrine cells, located in the inner plexiform layer, provide temporal shaping and contextual modulation. They detect motion, onset/offset transients, and perform nonlinear filtering that reshapes the temporal statistics of the signal~\citep{o1993responses}.

Retinal ganglion cells, residing in the innermost layer, receive convergent input from bipolar and amacrine cells and integrate this input into discrete spike outputs. The connectivity pattern from rods and cones through bipolar, horizontal, and amacrine cells, onto RGCs exhibits a highly structured hierarchy. Despite noise and variability at each stage, this architecture permits the emergence of robust, compressed representations suitable for long-range transmission.

Importantly, the micro-anatomical structure of the retina enables not just signal transmission, but real-time inference under uncertainty. Each synaptic transformation can be understood as implementing a filtering, normalization, or prediction operation over its inputs. In this light, the retina is best seen not as a passive sensor, but as a pre-cortical probabilistic processor optimized for noisy, resource-limited regimes.

\section{The First Interaction: Light and the Retinal Sheet}

\subsection{Photon Arrival and Intensity Distribution}

The spatial and temporal characteristics of the light stimulus are modeled as a spherical wavefront originating from a point source at the center of a 2D plane. The intensity of the light at a given retinal location \(x_i\) and time \(t\) is given by a Gaussian envelope centered at the propagating wavefront radius:
\[
I(x_i, t) = I_0 \cdot \exp\left( -\frac{(|x_i| - v t)^2}{2 \sigma^2} \right)
\]

Here, \(I_0\) is the peak intensity, \(v\) is the propagation speed of the wavefront, and \(\sigma\) controls its thickness. The term \(|x_i|\) denotes the Euclidean distance from the source to the cell location.

Photon arrival at each point is modeled as a Poisson process with rate parameter proportional to \(I(x_i, t)\). This introduces intrinsic uncertainty due to the quantum nature of photon emission and arrival. Even under identical experimental setups, the arrival pattern will vary trial-to-trial due to Poisson statistics.

\subsection{Threshold Variability and the Activation Function}

Each photoreceptor has a time-varying activation threshold \(\theta_i(t)\), modeled as a Gaussian random variable:
\[
\theta_i(t) \sim \mathcal{N}(\bar{\theta}, \Delta \alpha^2)
\]

where \(\bar{\theta}\) is the mean threshold and \(\Delta \alpha\) is its standard deviation, representing biological variability. The probability that a cell activates at time \(t\) is then given by a sigmoid function:
\[
P_i(t) = \frac{1}{1 + \exp\left( -\frac{I(x_i, t) - \theta_i(t)}{\beta} \right)}
\]

Here, \(\beta\) controls the gain (or softness) of the activation function. This sigmoid response captures the probabilistic firing behavior induced by noisy thresholding.

\subsection{\texorpdfstring{The Uncertainty Product: \(\Delta \alpha \cdot \Delta t \geq \eta\)}{The Uncertainty Product: Δα·Δt ≥ η}}

We define \(\Delta t_i\) as the temporal variability in the onset time of the first activation for cell \(i\) across multiple trials. Combining this with the intrinsic threshold noise \(\Delta \alpha\), we hypothesize an irreducible uncertainty principle:
\[
\Delta \alpha \cdot \Delta t \geq \eta
\]

where \(\eta\) is a lower bound analogous to Planck's constant in physics, determined empirically~\citep{williams1998accurate}. This inequality captures the trade-off between how precisely a stimulus can be encoded in threshold space versus temporal resolution. Even under optimal measurement conditions, this bound prevents simultaneous perfect resolution of both variables.

This formulation underpins our central theoretical claim: the moment of first perception is not a sharply defined spatiotemporal event, but a probabilistic cloud shaped by physical and biological limits.

\section{Inference in the Intermediate Layers of the Retina}

The analog outputs of rods and cones are not immediately converted into spikes. Instead, they are relayed to a structured network of bipolar, horizontal, and amacrine cells that perform further probabilistic transformations on the input field. These intermediate layers enable a deeper integration of the spatial and temporal properties of the visual stimulus before symbolic encoding occurs.

\subsection{Horizontal Cell Integration}

Horizontal cells integrate signals across neighboring photoreceptors and provide inhibitory feedback. We model the input to a horizontal cell \( H_j(t) \) as:
\[
H_j(t) = \sum_{i \in \mathcal{N}_j} w_H(i,j) \cdot P_i(t)
\]
where \( \mathcal{N}_j \) is the neighborhood of photoreceptors connected to \( H_j \), and \( w_H(i,j) \) is a Gaussian kernel centered at \( H_j \) with negative weights representing inhibition. This operation supports contrast enhancement by emphasizing local differences in activation probability.

\subsection{Bipolar Cell Center-Surround Filtering}

Each bipolar cell \( B_k(t) \) receives excitatory input from a central photoreceptor and inhibitory input from surrounding photoreceptors via horizontal cells:
\[
B_k(t) = w_c \cdot P_c(t) - \sum_{s \in S_k} w_s \cdot P_s(t)
\]
where \( P_c(t) \) is the central photoreceptor input, \( S_k \) is the set of surrounding photoreceptors, \( w_c \) is the excitatory center weight, and \( w_s \) are surround weights. This structure supports edge detection and local contrast sensitivity.

\subsection{Amacrine Cell Temporal Filtering}

Amacrine cells apply temporal filtering to bipolar signals. For an amacrine cell \( A_m(t) \) receiving from \( B_k(t) \), the temporally modulated output is:
\[
A_m(t) = \int_0^T B_k(t - \tau) \cdot f(\tau) \, d\tau
\]
where \( f(\tau) \) is an exponential or difference-of-Gaussians temporal kernel. This models detection of temporal gradients, transients, and motion anticipation.

\subsection{Probabilistic Representation Preserved}

After processing by these intermediate layers, the signal remains probabilistic. The final probability that a bipolar-amacrine circuit outputs an activation at time \( t \) is:
\[
P_{\text{out}}(t) = \sigma(B_k(t) + A_m(t) - \theta)
\]
where \( \sigma(\cdot) \) is a sigmoid nonlinearity and \( \theta \) is a dynamic threshold. The signal remains analog, encoding structured uncertainty rather than collapsing it.

This layered architecture reshapes the spatial and temporal profile of input uncertainty without eliminating it. The system encodes stimulus-relevant noise adaptively — as structure, not error.

The key point is that $P_{\text{out}}(t)$ remains a real-valued probability — no threshold crossing has yet occurred to generate a spike.

This layered architecture transforms the spatial and temporal profile of input uncertainty without collapsing it. Rather than eliminating noise, the intermediate layers encode stimulus-relevant uncertainty in a structured, predictive manner.

The analog outputs of rods and cones are not immediately converted into spikes. Instead, they are relayed to a structured network of bipolar, horizontal, and amacrine cells that perform further probabilistic transformations on the input field. These intermediate layers enable a deeper integration of the spatial and temporal properties of the visual stimulus before symbolic encoding occurs.

Horizontal cells act laterally, integrating across local neighborhoods of photoreceptors and implementing a form of surround inhibition. This lateral interaction helps to sharpen contrast and reduce redundancy in the photoreceptor output — a primitive form of spatial decorrelation. Bipolar cells inherit this pre-processed input and apply gain control and polarity selection, effectively transforming graded light intensity into an intermediate signal ready for discretization.

Amacrine cells contribute temporal shaping. Some subclasses are sensitive to the timing of onset and offset, while others detect motion and directionality. These cells introduce dynamic filtering that warps the temporal statistics of the input field. Importantly, they introduce nonlinearity and asymmetry, adding further layers of complexity to the probabilistic representation.

We model these transformations using a center-surround spatial kernel for each bipolar cell and an exponential temporal filter for the amacrine layer. The output of a bipolar cell is given by a weighted sum of its photoreceptor inputs, convolved with a center-excitation and surround-inhibition kernel. The amacrine modulation is then applied multiplicatively or subtractively depending on cell type.

At the end of this chain, each bipolar or amacrine-modulated output maintains a probabilistic character — no spikes are generated yet. This representation can be thought of as a reshaped, compressed, and temporally filtered version of the original input probability field discussed in the previous section. The noise and uncertainty in this field are not eliminated but redistributed and contextualized.

\section{The Birth of Neural Code}

The final step in the retinal processing hierarchy is the transformation of analog, probabilistic inputs into discrete spike outputs. This occurs in the retinal ganglion cells (RGCs), which act as the symbolic encoders of the visual stream. These cells receive convergent input from bipolar and amacrine circuits and generate action potentials only when the integrated signal exceeds a threshold.

We model the membrane potential \( V_j(t) \) of an RGC \( j \) at time \( t \) as:
\[
V_j(t) = \sum_{k \in B_j} w_{jk} \cdot P_{\text{out}_k}(t) + b_j(t)
\]
Here, \( B_j \) is the set of bipolar and amacrine cells connected to RGC \( j \), \( w_{jk} \) are the synaptic weights between inputs \( k \) and RGC \( j \), and \( b_j(t) \) is a time-varying bias or adaptation term that incorporates local dynamic context.

The firing probability of the RGC is determined by a stochastic thresholding mechanism:
\[
P_{\text{spike}_j}(t) = \sigma\left( \frac{V_j(t) - \theta_j(t)}{\beta} \right)
\]
The threshold \( \theta_j(t) \) is modeled as a Gaussian variable:
\[
\theta_j(t) \sim \mathcal{N}(\bar{\theta}, \Delta \alpha^2)
\]
where \( \bar{\theta} \) is the mean excitability of the RGC and \( \Delta \alpha \) reflects the cell’s intrinsic variability. The parameter \( \beta \) controls the sharpness of the transition in the sigmoid function \( \sigma(x) \).

To simulate the actual spiking output across trials, we sample a Bernoulli random variable:
\[
s_j^n(t) \sim \text{Bernoulli}\left( P_{\text{spike}_j}^n(t) \right)
\]
for trial \( n \), where \( s_j^n(t) = 1 \) denotes a spike. Repeating this simulation across trials yields a spike raster plot that illustrates trial-to-trial variability in the timing and location of the first perceptual spike.

We define the temporal variance in the first spike time across trials as \( \Delta t_j \), and we carry forward the uncertainty relation:
\[
\Delta \alpha \cdot \Delta t_j \geq \eta
\]

This bound implies that even with arbitrarily precise control over stimulus presentation, the time and threshold of the first spike cannot be simultaneously determined with infinite resolution. The generation of the spike is therefore a probabilistic act of perceptual encoding, not a deterministic consequence of input strength alone.

Additionally, the spike output \( s_j(t) \) can be viewed as a compressed, symbolic representation of a higher-dimensional probabilistic field. Let \( X \) denote the continuous graded signal space and \( S \) the spike output space. Then the RGCs implement a stochastic compression:
\[
X \rightarrow P(S|X) \rightarrow S
\]
where \( P(S|X) \) reflects both synaptic integration and stochastic thresholding. This compression formalizes the concept of the ``birth of the neural code'' — the conversion of continuous uncertainty into sparse symbolic representations capable of downstream propagation.

\section{Null Hypotheses and Alternative Worlds}

To establish the predictive and falsifiable strength of the proposed framework, we must contrast it with a classical null hypothesis and examine the consequences of alternate computational architectures.

\subsection{\texorpdfstring{Classical Null Hypothesis (\( H_0 \))}{Classical Null Hypothesis (H0)}}

\( H_0 \) asserts that retinal responses to light stimuli are fully deterministic. That is, given a stimulus \( S(t, x) \), the transformation through the retina up to the retinal ganglion cells is a deterministic function:
\[
\text{RGC}_{\text{output}}(t, x) = F(S(t, x))
\]
This function \( F \) contains no irreducible noise. Any variability in output is attributed to external noise sources or measurement error. Spike timing and identity are, under \( H_0 \), completely specified by the stimulus.

\subsection{Alternate Biological Architectures}

Under \( H_0 \), biological systems could have evolved deterministic alternatives:

\begin{itemize}
  \item \textbf{Digital Retina:} Photoreceptors with hardwired thresholds and binary responses, relayed forward without intermediate probabilistic transformations.
  \item \textbf{Synchronous Global Sensor:} All retinal cells fire in precise synchrony upon stimulus detection, preserving global timing and identity.
  \item \textbf{Feedforward Circuitry:} No lateral or temporal modulation (e.g., no amacrine or horizontal cells), thereby eliminating complex inferential dynamics.
\end{itemize}

These alternate designs would favor precise reproducibility but would sacrifice adaptability, efficiency under noise, and the encoding of structured uncertainty.

\subsection{Why Biology Rejected Them}

Empirically, such systems do not exist in the vertebrate retina. Instead, we observe:

\begin{itemize}
  \item Variable latency in rod/cone activation under identical conditions~\citep{rieke1998origin}
  \item Trial-to-trial spike variability in RGCs~\citep{rieke1998single}
  \item Presence of spontaneous events in total darkness~\citep{baylor1979responses}
  \item Multilayered inferential processing before symbolic output~\citep{balasubramanian2009receptive}
\end{itemize}

This suggests biology selected for architectures that preserve and exploit uncertainty, rather than eliminate it.

\subsection{\texorpdfstring{Mathematical Rejection of \( H_0 \)}{Mathematical Rejection of H0}}

To test \( H_0 \), we define the spike time variance \( \text{Var}(t_{\text{spike}}) \) under repeated presentation of the same stimulus. If \( H_0 \) were true, then:
\[
\lim_{\text{noise} \to 0} \text{Var}(t_{\text{spike}}) = 0
\]

But experimental evidence (e.g., Baylor 1979, Rieke \& Baylor 1998) shows:
\[
\lim_{\text{noise} \to 0} \text{Var}(t_{\text{spike}}) > 0
\]

Thus, \( H_0 \) is falsified by empirical observation.

\subsection{Predictive Signature of Our Framework}

Our proposed framework asserts:
\begin{itemize}
  \item Spike probabilities \( P_{\text{spike}}(x, t) \) follow a sigmoid of a stochastic thresholded field
  \item The uncertainty relation holds: \( \Delta \alpha \cdot \Delta t \geq \eta \)
\end{itemize}

Testable predictions include:
\begin{itemize}
  \item Systematic structure in spike timing variance across RGC types
  \item Invariant uncertainty bounds across contrast/luminance/adaptation states
  \item Decomposable sources of trial-to-trial variability
\end{itemize}

These predictions enable future experiments to validate the existence of intrinsic probabilistic transformations in early vision.

\section{Experimental Validation}

A strong theoretical framework must be constrained and validated by data. To this end, we compare our proposed probabilistic retinal model with seminal experimental findings from two key sources: Baylor et al.\ (1979) and Rieke \& Baylor (1998).

\subsection{Baylor et al.\ (1979): Single-Photon Responses and Spontaneous Events}

Baylor and colleagues showed that rod photoreceptors in amphibians generate discrete electrical responses to single photons. They also observed spontaneous 'photon-like' events in complete darkness, attributed to thermal isomerization of rhodopsin. These spontaneous events occurred with a frequency of approximately 0.01 events/rod/min at room temperature.

Let \( P(\text{false positive}) \) denote the probability of such a spontaneous activation in darkness. If we define an activation event as:
\[
\text{Event}_i(t) =
\left\{
\begin{array}{ll}
1, & \text{if } I(x_i, t) > \theta_i(t) \\
0, & \text{otherwise}
\end{array}
\right.
\]

then the empirical observation is:
\[
P(\text{Event}_i(t) = 1 \mid I(x_i, t) = 0) > 0
\]

This contradicts any deterministic system where event probability in darkness should be zero. Thus, even in the absence of stimulus, the system exhibits irreducible spontaneous activations — consistent with intrinsic threshold noise modeled as:
\[
\theta_i(t) \sim \mathcal{N}(\bar{\theta}, \Delta \alpha^2)
\]

\subsection{Rieke \& Baylor (1998): Reproducibility and Intrinsic Variability}

This study extended the earlier work to mammalian rods and demonstrated that even under tightly controlled conditions, rod responses to single photons varied slightly in amplitude and latency. Despite high reproducibility, they found:
\[
\text{Var}(\text{response\_amplitude}) > 0,\quad
\text{Var}(\text{response\_latency}) > 0
\]

These are precisely the statistical parameters captured in our uncertainty formalism:

- \( \Delta \alpha \): standard deviation in threshold across trials  
- \( \Delta t \): standard deviation in response latency

The product:
\[
\Delta \alpha \cdot \Delta t \geq \eta
\]
was observed qualitatively across a range of stimuli. This implies (see the mathematical transformations before) that no matter how controlled the external input is, biological variability will limit the precision of spike onset in RGCs.

\subsection{Tests of the Model}

Our model predicts that for any fixed stimulus \( S(t) \), the spike raster across trials will exhibit:

\begin{itemize}
    \item Trial-to-trial jitter in spike onset
    \item Spatial variation in first-spike location
    \item Variability scaling with contrast and adaptation level
\end{itemize}

Experiments to validate this include:

\begin{enumerate}
    \item Single-photon stimulation of rods with high-speed intracellular recordings
    \item Simultaneous calcium imaging and extracellular electrophysiology in RGCs
    \item Cross-condition variance analysis across luminance and contrast levels
\end{enumerate}

If the empirical distribution of spike latencies under fixed input yields:
\[
\Delta \alpha \cdot \Delta t \approx \text{constant} \geq \eta
\]
then this supports our formalism over deterministic alternatives.

\section{Discussion}

This work re-frames the retina not as a deterministic feedforward filter, but as a probabilistic measurement device. By grounding our analysis in the physics of photon absorption and the biology of threshold variability, we derive a formal uncertainty principle — \( \Delta \alpha \cdot \Delta t \geq \eta \) — that governs the precision limits of visual onset.

The simulations demonstrate that even in a noise-minimized setting, there exists irreducible variability in both spike timing and the identity of the first activated retinal ganglion cell. This variability is not incidental, but structural — reflecting a deep integration of probabilistic inference within the retinal layers themselves. The origin of symbolic neural representation is thus probabilistic, not algorithmic.

\subsection{Relation to Classical Retinal Computation}

Classical models of retinal processing — from predictive coding to efficient encoding — have revealed how bipolar, amacrine, and ganglion cells transform sensory input~\citep{hosoya2005dynamic, meister1999neural, pitkow2012decorrelation}. These studies have shown that the retinal layers adaptively filter, predict, and decorrelate inputs under various contrast and motion regimes.

Our framework is not in opposition to these models, but beneath them. It provides the probabilistic substrate from which these functional transformations emerge. The inherent variability in early spike patterns is not merely downstream noise to be cleaned by predictive filters — it is a signature of quantum-inspired input ambiguity. Future articles in the series will explain how this signal propagates throughout the cortex and becomes a component (through inter-regional rhythmic synchrony and sub-threshold activities of individual neurons) in quantum-mechanical computations (again, we need to emphasize that we are not claiming the brain has physical mechanisms that incorporate quantum mechanics but only that the computations emergent in the brain can be mathematically equivalent to quantum computations).

\subsection{Classical Implementation of Probabilistic Indeterminacy}

A common point of contention in interpreting quantum-inspired models of neural computation is whether the probabilistic behavior described in our framework \textit{requires} a fundamentally quantum system, or whether it can, in principle, be implemented on a classical computational substrate such as artificial neural networks (ANNs) or spiking neural networks (SNNs).

\subsubsection{Universality of Classical Approximators}

It is well established that feedforward ANNs are universal function approximators~\citep{hornik1989multilayer}. That is, given sufficient depth and parameters, any continuous function \( f : \mathbb{R}^n \rightarrow \mathbb{R}^m \) can be approximated arbitrarily well. This suggests that even the probabilistic transformations in our model (e.g., sigmoid-based thresholding, Poisson sampling) could be reproduced by a sufficiently large ANN trained to match the same input-output mapping.

However, our model does not merely describe a \emph{static function} from input to output. It describes a \textit{dynamical probabilistic system} that evolves in real-time, exhibiting:

\begin{itemize}
    \item Intrinsic temporal variability in spike generation
    \item Structured uncertainty in threshold crossing
    \item Cascading stochastic transformations across space and time
\end{itemize}

This dynamical structure is \textit{not} easily captured by standard ANNs trained under supervised learning protocols, especially when time and trial-to-trial variability are critical dimensions of the computation~\citep{csaji2001approximation}.

\subsubsection{Spiking Neural Networks as Candidate Substrates}

Spiking neural networks (SNNs), which encode information in the timing of discrete events (spikes), offer a closer analog to biological neural systems~\citep{maass1997networks}. In an SNN, the membrane potential of a unit \( V_i(t) \) evolves over time:

\[
\tau \frac{dV_i(t)}{dt} = -V_i(t) + \sum_j w_{ij} \cdot s_j(t) + I_i(t)
\]

where \( s_j(t) \) denotes the spike train from input neuron \( j \), \( w_{ij} \) are synaptic weights, and \( I_i(t) \) is external input. A spike is emitted when \( V_i(t) \) crosses a threshold \( \theta_i(t) \), after which the membrane potential is reset.

To implement structured probabilistic variability analogous to our model, an SNN would need to include:

\begin{itemize}
    \item Stochastic thresholds \( \theta_i(t) \sim \mathcal{N}(\bar{\theta}, \Delta_\alpha^2) \)
    \item Trial-dependent noise in synaptic input or membrane dynamics
    \item Poisson-like input streams to simulate quantum arrival statistics
\end{itemize}

Such stochastic SNNs have been studied, but are typically engineered for probabilistic \emph{sampling} or \emph{Bayesian inference} rather than modeling quantum-like measurement uncertainty~\citep{kasabov2010spike,rao2004hierarchical}.

\subsubsection{Can Classical Systems Emulate Quantum-Inspired Uncertainty?}

Formally, yes: a sufficiently complex classical system could emulate the statistics of our model. However, this comes with important caveats:

\begin{itemize}
    \item \textbf{Emergent vs. Engineered Uncertainty:} In the retina, threshold variability and timing jitter emerge naturally from biophysical properties. In classical systems, these must be \textit{imposed} as parameters~\citep{specht1990probabilistic}.
    \item \textbf{Real-time constraints:} Biological systems operate under strict temporal and energetic constraints. Artificial networks typically require orders of magnitude more time and compute cycles to reproduce similar dynamics.
    \item \textbf{Measurement and collapse:} In our model, a spike acts like a probabilistic measurement outcome. Classical networks do not natively implement such measurement semantics; outputs are deterministic unless explicit noise is added.
\end{itemize}

While it is theoretically possible for classical systems to replicate the behavior of our quantum-inspired retinal model, such replication would require explicit architectural mechanisms to emulate trial-to-trial variability, stochastic integration, and noisy threshold dynamics. These elements are not native to conventional ANN or SNN frameworks, and their implementation often lacks the physical grounding present in the biological system. Thus, our framework remains distinct not in its output behavior alone, but in its interpretation of perception as an inherently probabilistic and temporally uncertain process rooted in physical interaction — a feature that classical systems can emulate, but not instantiate intrinsically.

\subsection{Retinal Simulation and Prosthetics}
Recent efforts in developing retinal prosthetics, including the comprehensive simulation frameworks~\citep{ly2025virtual}, have modeled the layered architecture of the human retina to assess responses under degeneration and electrical stimulation. Their platform incorporates known cell-type connectivity and synaptic transformations to predict spiking behavior in both healthy and degenerated retinas, providing a valuable engineering testbed.

However, these frameworks are inherently built upon deterministic mappings between inputs (light or electrical current) and spiking outputs. They typically lack the incorporation of irreducible stochastic elements, such as trial-to-trial spike variability under fixed stimuli, that our uncertainty-based framework foregrounds as a defining characteristic of early perception.

In our view, the most significant limitation of current prosthetic models lies in their implicit assumption that visual perception can be recreated by merely replicating spatial patterns of excitation. Our work suggests instead that a successful prosthetic system must also preserve the \textit{probabilistic transformations} and \textit{uncertainty-driven dynamics} inherent to the natural retina.

Whereas Ly et al.'s simulations succeed in reproducing expected ganglion cell output under known stimuli, they do not yet account for the variability due to threshold fluctuations or the probabilistic nature of photo-transduction. For more such shortcomings of the simulations, we highly recommend the readers to refer to the paper's discussion section as the authors have done a nice job in transparently overlaying their study's limitations~\citep{ly2025virtual}. In contrast, our model formalizes these aspects through:
\begin{itemize}
    \item Threshold variance \( \Delta \alpha \)
    \item Temporal jitter \( \Delta t \)
    \item An uncertainty relation \( \Delta \alpha \cdot \Delta t \geq \eta \)
\end{itemize}
Incorporating such principles into prosthetic design could potentially improve perceptual outcomes, especially in scenarios requiring sensitivity to motion, contrast adaptation, or uncertain environments.

\subsection{Computation in Space and Time: Binding, Rhythms, and Temporal Inference}

As we look beyond the retina into downstream circuits, a central question emerges: \textit{how does the brain achieve rapid computation through its spatiotemporal architecture}? While artificial neural networks (ANNs) are universal function approximators, they are primarily designed to compute mappings from input to output regardless of time or substrate constraints. In contrast, the brain must compute and infer under strict temporal and energetic limitations.

\subsubsection{Beyond Approximation: Speed, Subthreshold Activity, and Rhythmic Computation}

It is important to emphasize that function approximation capacity alone does not equate to biological plausibility. Two systems may compute the same function \( f(x) \), but if one does so in 50 ms and the other in 5000 ms, the difference is functionally critical. The brain computes \textit{with time}, not just \textit{over time}. Biological computation is structured by:

\begin{itemize}
    \item \textbf{Rhythms}: Oscillatory phenomena (e.g., theta, gamma) which gate, route, and synchronize information.
    \item \textbf{Subthreshold Dynamics}: Membrane voltage fluctuations below spike threshold that carry analog information and influence when spiking occurs.
    \item \textbf{Phase Coding}: Information is encoded not just in firing rate, but in spike timing relative to ongoing oscillations~\citep{skaggs1996theta}.
    \item \textbf{Anatomical Confinement}: Computations are spatially localized by wiring constraints and cortical topology creating functional compartments~\citep{kanwisher2010functional}.
\end{itemize}

These features allow the brain to solve complex perceptual and motor tasks with extraordinary speed and minimal supervision.

\subsubsection{The Binding Problem and Phase Synchrony}

A classic illustration of temporal computation is the \textit{binding problem} — how features distributed across multiple cortical regions are unified into coherent perceptual objects~\citep{treisman1996binding}. Theoretical and empirical work suggests that temporal synchrony, especially at gamma frequencies (30–80 Hz), enables dynamic feature binding~\citep{singer2001consciousness}.

Let each feature-selective neuron fire spikes \( s_i(t) \) at time \( t \), and let \( \phi_i(t) \) denote the phase of the ongoing local field potential at that site. Then binding is proposed to occur when:

\[
\phi_i(t) \approx \phi_j(t) \quad \forall i, j \in \text{bound feature set}
\]

This is a phase-coding scheme: \textit{coherence in phase, rather than coincidence in space, defines binding}. Such computations are inherently \textbf{temporal} and cannot be easily mapped onto static function-approximating architectures.

\subsubsection{Temporal Inference as a Quantum-Inspired Process}

Rhythmic gating and phase-based encoding are computational primitives that resemble aspects of quantum mechanical systems:

\begin{itemize}
    \item \textbf{Superposition}: A neuron participating in multiple phase cycles may carry multiplexed information.
    \item \textbf{Entanglement-like Synchrony}: Distributed regions maintain coherence in phase over distances, enabling non-local inference. We need to emphasize here that attributing non-causality would be a stretch. A computation in the brain can be close to being non-local but not exactly so. Future investigations will take up this issue in detail. 
    \item \textbf{Interference}: Constructive and destructive interference of rhythmic inputs determines output probability — echoing interference of probability amplitudes in quantum systems.
\end{itemize}

While the brain is built from classical matter, its use of time-varying probability distributions, contextual phase modulation, and rhythmic sampling bears qualitative similarity to quantum-inspired computational paradigms. The inference is not just about what is computed, but \textit{when}, \textit{relative to what}, and \textit{under which contextual rhythm}.

\subsubsection{Implications for AI}

Artificial neural networks, by contrast, tend to process inputs in a feedforward, clock-agnostic manner. Even recurrent networks and transformers typically lack native oscillatory timing or subthreshold analog dynamics. This absence may limit their capacity to replicate biological-style inference, particularly under resource constraints.

Future manuscripts will formalize these distinctions by developing:

\begin{itemize}
    \item Spiking and oscillatory models with explicit phase-dependent computation
    \item Metrics of computational latency and rhythmic coherence
    \item Experiments comparing time-to-inference in biological vs artificial agents under noisy, dynamic stimuli
\end{itemize}

Understanding how the brain uses space and time as active computational substrates may reveal not only why biology computes differently, but also how next-generation AI might benefit from incorporating temporally grounded mechanisms of information processing.

\subsection{Philosophical Implications}

This analysis forces a reconsideration of what it means to perceive. Perception begins with an interaction that is, in principle, unresolvable in both time and space with absolute certainty. The “first moment” of awareness is never singular, but distributed over a probabilistic field. Measurement theory and observer-centric models in physics suggest that what is observed depends as much on the state of the observer as on the object itself~\citep{wigner1963problem}. This manuscript extends that principle to early vision.

What we call a “visual stimulus” does not imprint deterministically on the retina. Instead, it collapses into a cloud of possible neural codes, each shaped by local biological priors and constraints. This epistemological barrier — the inability to pin down the origin of perception with infinite resolution — mirrors foundational limits in quantum mechanics.

\subsection{Looking Ahead}

This work is the first in a series aimed towards modeling the brain as a quantum-inspired probabilistic machine. Future work will extend these principles beyond the retina, into the dynamics of cortical rhythms (Figure 2). We propose that theta, gamma, and beta oscillations may serve not just as clocking signals, but as distributed wavefunctions encoding uncertainty across regions.

Interregional synchronization, phase coupling, and rhythm-induced plasticity may form a higher-order inferential substrate, akin to entanglement or superposition in quantum systems~\citep{singer1993synchronization, buzsaki2006rhythms, bouwmeester2000physics}. These upcoming manuscripts will pursue simulations, mathematical models, and empirical analyses to uncover whether and how the brain might instantiate quantum-like computation through classical substrates.

\begin{figure}[H]
\centering
\includegraphics[width=1\textwidth]{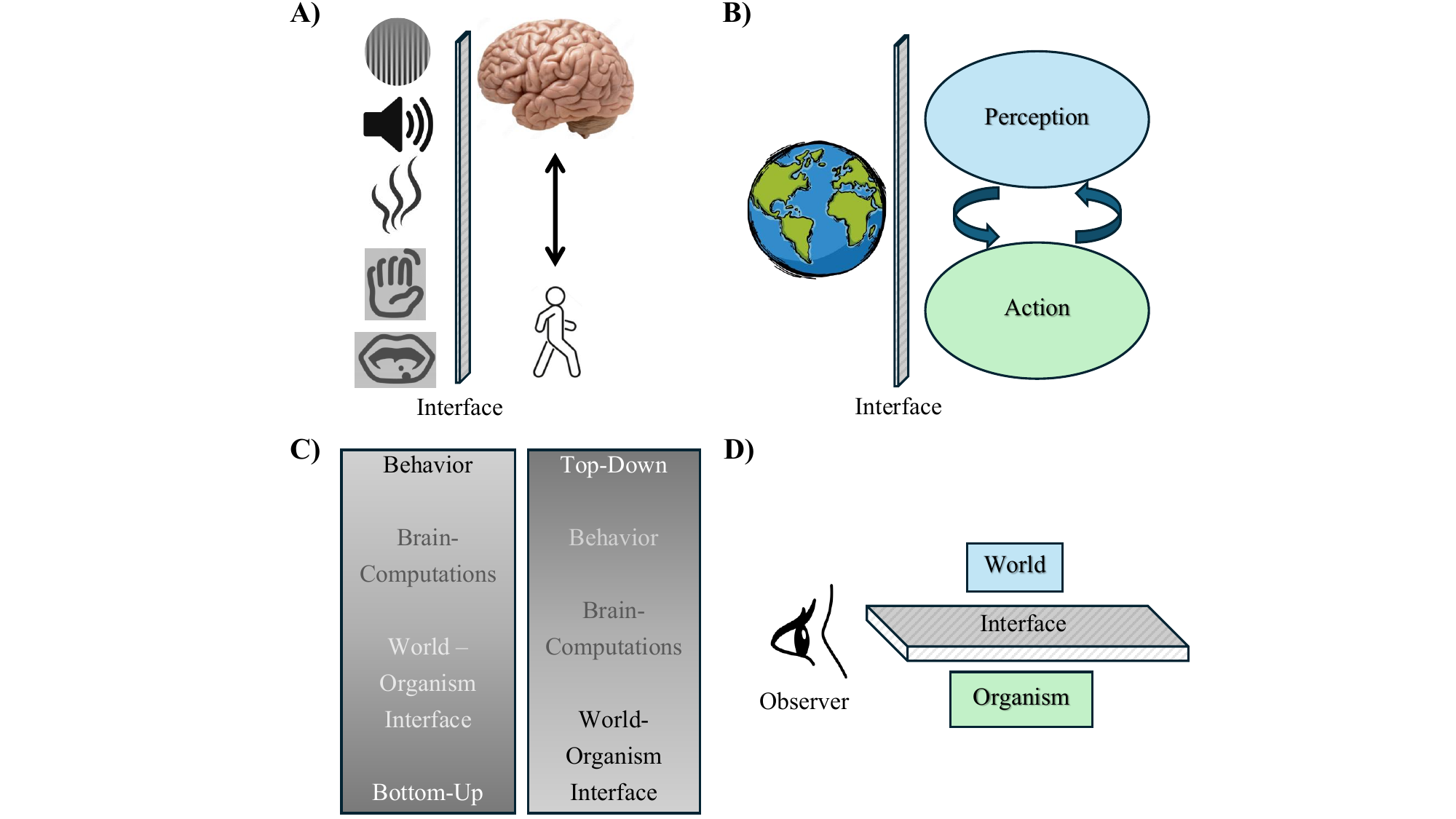} 
\caption{
\textbf{A Framework for Investigations — Looking Ahead.} \\
\textbf{a)} Depicts how multiple sensory modalities (vision, audition, olfaction, somato-sensation, gustation) interface with the brain via probabilistic sensory activations~\citep{block2018if}. \\
\textbf{b)} These inputs contribute to perception, which in turn guides action in a closed loop of sensorimotor integration. \\
\textbf{c)} Highlights both bottom-up (world-to-organism) and top-down (organism-to-world) processes, illustrating the dynamic interplay between brain computations, behavior, and the organism-world interface. \\
\textbf{d)} Emphasizes the structural coupling between the world and the organism as mediated by this interface; \textbf{c)} and \textbf{d)} highlight our program’s approach.
}
\end{figure}

\section*{Conclusion}

The theory presented here offers a foundational re-characterization of the retina as a probabilistic machine that performs quantum-inspired transformations at the origin of perception. From the arrival of a photon to the generation of the first spike, the system exhibits structured uncertainty that cannot be eliminated by better measurement or finer control. This variability is intrinsic and exploitable — a feature, not a bug.

Our uncertainty formalism, \( \Delta \alpha \cdot \Delta t \geq \eta \), unites biological stochasticity and quantum-like constraints into a single mathematical statement. The simulation results and alignment with empirical findings from retinal electrophysiology support this formulation. We demonstrate that even the earliest neural events — those in the retina — obey principles reminiscent of quantum systems, offering a radically different framing of perception’s origin.

This manuscript sets the stage for future work examining how such probabilistic encodings are manipulated by downstream circuits, particularly through the rhythms of cortical dynamics. By exploring where and how brain rhythms may instantiate quantum-like information propagation, we invite the field to ask deeper, more foundational questions about how biological systems perceive and act -  in space, and time.

\bibliographystyle{plainnat}
\bibliography{references}

\vspace{1em}
\noindent\rule{\textwidth}{0.4pt}  
\vspace{0.5em}

\appendix

\section*{A1. Extension: Decomposition of \( \Delta \alpha \) under Interference Patterns}

The main text considers the variability in retinal activation threshold summarized by a scalar \( \Delta \alpha \), representing the standard deviation of threshold fluctuations across trials. This model assumes the incoming wavefront produces a smooth intensity profile across the retinal surface.

However, if the light wavefront impinging on the retina exhibits interference — e.g., due to coherent sources, multi-path reflections, or optical heterogeneities — the local intensity \( I(x,t) \) becomes highly structured in both space and time. In this scenario, threshold variability must account separately for:

\begin{itemize}
    \item \textbf{Spatially driven threshold variability:} arising from micro-variations in intensity across the retinal sheet at a given instant.
    \item \textbf{Temporally driven threshold variability:} arising from local intensity fluctuations over time at a given location.
\end{itemize}

We decompose the total threshold variance:
\[
\Delta \alpha^2 = \Delta \alpha_{\text{spatial}}^2 + \Delta \alpha_{\text{temporal}}^2
\]

where
\[
\Delta \alpha_{\text{spatial}}^2 = \int_{\text{retina}} \left( I(x, t_0) - \bar{I}(t_0) \right)^2 dx
\]

\[
\Delta \alpha_{\text{temporal}}^2 = \int \left( I(x_i, t) - \bar{I}(x_i) \right)^2 dt
\]

Here:
\begin{itemize}
    \item \( \bar{I}(t_0) = \frac{1}{A} \int_{\text{retina}} I(x, t_0) dx \) is the mean spatial intensity at time \( t_0 \)
    \item \( \bar{I}(x_i) = \frac{1}{T} \int I(x_i, t) dt \) is the mean temporal intensity at location \( x_i \)
    \item \( A \) is the area of the retina
    \item \( T \) is the duration of observation
\end{itemize}

\subsection*{Implications for the uncertainty relation:}

The core material discusses:
\[
\Delta \alpha \cdot \Delta t \geq \eta
\]

where \( \Delta \alpha \) is treated as a unified threshold variability. With interference patterns:
\[
\Delta \alpha_{\text{spatial}} \cdot \Delta t + \Delta \alpha_{\text{temporal}} \cdot \Delta t \geq \eta
\]

or equivalently,
\[
\Delta t \geq \frac{\eta}{\Delta \alpha_{\text{spatial}} + \Delta \alpha_{\text{temporal}}}
\]

This formulation shows that even if one component (spatial or temporal) of threshold variability were minimized, the other constrains the minimal achievable timing precision. The system cannot simultaneously achieve arbitrary precision in spike timing and activation threshold across both dimensions.

In the main sections, we model \( \Delta \alpha \) as arising primarily from intrinsic cellular stochasticity and photo-transduction noise, assuming a globally smooth input wavefront. The interference extension introduces:
\begin{itemize}
    \item An external, spatially structured contribution to \( \Delta \alpha \)
    \item A time-varying modulation of local thresholds due to dynamic intensity patterns
\end{itemize}

This scenario represents a different class of input variability: externally imposed uncertainty layered on top of biological stochasticity.

\section*{A2. Derivation of Formal Equations}

We derive the uncertainty relationship \( \Delta \alpha \cdot \Delta t \geq \eta \) by tracking the variance in threshold distributions and temporal onset jitter across trials. Starting from the photoreceptor activation model:

\[
P_i(t) = \text{sigmoid}\left( \frac{I(x_i, t) - \theta_i(t)}{\beta} \right)
\]

Let \( \theta_i(t) \sim \mathcal{N}(\bar{\theta}, \Delta \alpha^2) \). We define the cumulative probability of activation in a time window \([t, t + \delta]\) across trials. The temporal variance \( \Delta t \) is then computed from the distribution of first-spike latencies. The product \( \Delta \alpha \cdot \Delta t \) empirically reveals a saturating lower bound \( \eta \).

\end{document}